\documentclass[reprint,prb,amsmath,amssymb,superscriptaddress,floatfix]{revtex4-1}
\usepackage{hyperref}
\usepackage{graphics}
\usepackage{enumerate}

\RequirePackage{lineno} 

\usepackage{color}

\usepackage{ulem}

\begin{document}

\title{Spatially resolving valley quantum interference of a donor in silicon}
\author{J. Salfi}
\author{J. A. Mol}
\affiliation{Centre for Quantum Computation and Communication Technology, School of Physics, The University of New South Wales, Sydney, NSW 2052, Australia.}
\author{R. Rahman}
\author{G. Klimeck}
\affiliation{Purdue University, West Lafayette, IN 47906, USA.}
\author{M. Y. Simmons}
\affiliation{Centre for Quantum Computation and Communication Technology, School of Physics, The University of New South Wales, Sydney, NSW 2052, Australia.}
\author{L. C. L Hollenberg}
\affiliation{Centre for Quantum Computation and Communication Technology, School of Physics, University of Melbourne, Parkville, VIC 3010, Australia.}
\author{S. Rogge}
\email{s.rogge@unsw.edu.au}
\affiliation{Centre for Quantum Computation and Communication Technology, School of Physics, The University of New South Wales, Sydney, NSW 2052, Australia.}
\begin{abstract}

Electron and nuclear spins of donor ensembles in isotopically pure silicon experience a vacuum-like environment, giving them extraordinary coherence.  However, in contrast to a real vacuum, electrons in silicon occupy quantum superpositions of valleys in momentum space.  Addressable single-qubit and two-qubit operations in silicon require that qubits are placed near interfaces, modifying the valley degrees of freedom associated with these quantum superpositions and strongly influencing qubit relaxation and exchange processes.  Yet to date, spectroscopic measurements only indirectly probe wavefunctions, preventing direct experimental access to valley population, donor position, and environment.  Here we directly probe the probability density of single quantum states of individual subsurface donors, in real space and reciprocal space, using scanning tunneling spectroscopy.  We directly observe quantum mechanical valley interference patterns associated with linear superpositions of valleys in the donor ground state.  The valley population is found to be within $5 \%$ of a bulk donor when $2.85\pm0.45$ nm from the interface, indicating that valley perturbation-induced enhancement of spin relaxation will be negligible for depths $>3$ nm.  The observed valley interference will render two-qubit exchange gates sensitive to atomic-scale variations in positions of subsurface donors.  Moreover, these results will also be of interest to emerging schemes proposing to encode information directly in valley polarization.  

\end{abstract}

\date{\today}

\maketitle

Fabrication of devices\cite{Morello:2010ga,Fuechsle:2012bl}, spin readout\cite{Morello:2010ga}, and quantum control of spins\cite{Pla:2012jj,Pla:2013ta} in silicon has been accomplished at the single-donor level.  However, addressable control and coupling within qubit arrays requires local gates and control interfaces, whose atomic-scale potentials strongly influence electronic valley degrees of freedom\cite{Kane:1998ce,Koiller:2001gw,Wellard:2005hs,Goswami:2006bb,Calderon:2006ha,Rahman:2007ev,Lansbergen:2008bs,Saraiva:2009gh,Morton:2011ek,Yang:2013if,Roche:2012gb,Zwanenburg:2013gl}.  While these unconventional orbital degrees of freedom play no role in conventional silicon microelectronics, they invariably arise in quantized states in indirect gap materials, and are pervasive in quantum electronics.  In silicon, valley physics determines qubit relaxation rates\cite{Roth:1960co,Hasegawa:1960ey,Morton:2011ek} and are predicted to strongly influence two qubit gates\cite{Koiller:2001gw,Wellard:2005hs}.  Moreover, valley degrees of freedom in AlAs\cite{Gunawan:2006bo}, silicon\cite{Takashina:2011jz}, diamond\cite{Isberg:2013hz}, and graphene\cite{Young:2012bn} can play a role similar to spin in condensed matter systems.  Encoding of  information within valley polarization has been proposed in AlAs\cite{Gunawan:2006hb}, silicon\cite{Soykal:2011dd,Culcer:2012bj} and diamond\cite{Isberg:2013hz}, and within polarization of chiral valley pseudospin in carbon-based nanostructures\cite{Rycerz:2007kd,Tombros:2011dh,Pei:2012jn}. 

Here, individual states of subsurface donors were measured using cryogenic scanning tunneling spectroscopy.  Quantum mechanical valley interference patterns were observed in real space, associated with linear quantum superpositions of wavevectors in the six conduction band valleys of silicon.  Enabled by high-accuracy empirical determination of donor depth and electric field, we perform a parameter-free comparison with atomistic theory establishing that the z-valley population is $\sim 38 \pm 2 \%$ for a  $2.85\pm 0.45$ nm deep donor, perturbed by only $\sim 5\%$ compared with $\sim 33.3 \%$ for a donor in bulk silicon.  Consequently, donors more than $3$ nm deep should not experience a significant valley-repopulation induced increase in spin-lattice relaxation.  Moreover, the nearly bulk-like valley interference observed will render two-qubit exchange gates sensitive to atomic variations in donor position\cite{Koiller:2001gw,Wellard:2005hs}, even for subsurface donors. Measurements of valley quantum interference presented herein address spatial aspects of the now 60-year old theory of shallow impurities in silicon\cite{Kohn:1955kc,Pantelides:1974ew,Wellard:2005hs,Rahman:2007ev} that are essential for engineering spin qubit arrays leveraging silicon's exceptional coherence\cite{Tyryshkin:2011fi,Steger:2012ev}. 

\begin{figure*}
\includegraphics{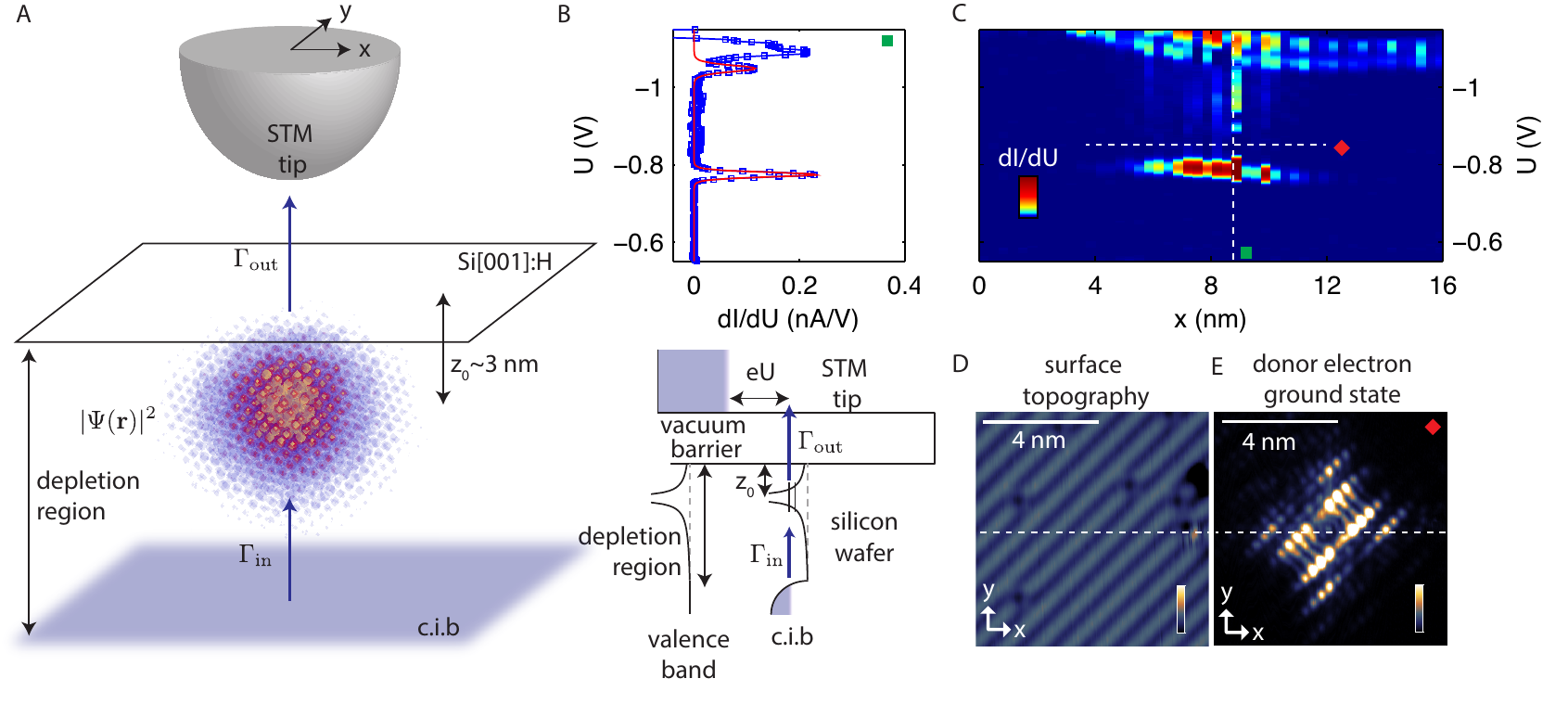}
\caption{\textbf{Spatial measurement of single quantum states of a subsurface donor} A.  Illustration of single-electron tunneling from the conduction impurity band (c.i.b.)\cite{VanMieghem:1992kx}, to the subsurface donor in the depletion region with electron probability density $|\Psi(\mathbf{r})|^2$, to the tip.  B.  Top: Measured $dI/dU$ (blue squares) for tip position above the subsurface donor, and least-squares fit to thermally broadened single-electron transport (red line).  Bottom: Schematic band diagram, whereby the sample bias $U<0$ brings the donor-bound states into resonance with the Fermi energy in the conduction impurity band (c.i.b.), which is spatially separated from the donor by a depletion region.  C.  Measured differential conductance $dI/dU$ versus $x$ and sample bias $U$, along line passing over single donor.  Scale: (0,0.25) nA/V.  Tip location for measurement in B is indicated by the vertical dashed line (green box).  D. Constant current ($I_0=150$ pA) topography $z(x,y)$ recorded during first line scan at $U=-1.45$ V, a bias where there is minimal evidence of buried donor. Scale: (0,100) pm.  E. Spatially resolved single-electron tunneling current $I(\mathbf{r})\propto|\Psi(\mathbf{r})|^2$ through the single donor-bound state, for the same area as D, but with $U=-0.85$ V, and the feedback loop off.  The signal vanishes away from the donor, as expected for a true single state measurement.  White dashed line (red diamond) denotes where data in 1C was obtained.  Scale: (0,0.35) nA.  }
\end{figure*}

We employed unconventional scanning tunneling spectroscopy (STS)-based spatially-resolved single-electron transport to measure, in real space, valley interference in single quantum states of subsurface arsenic donors.  Ultra-high vacuum annealing, with parameters targeted to deplete the upper $\sim 10$ nm of donors\cite{Pitters:2012gt}, was performed on a highly arsenic-doped silicon wafer (see Methods).  Spatially well-isolated residual donors (density $\sim 10^{11}$ cm$^{-2}$) were found in the depletion region by subsurface imaging\cite{Koenraad:2011ed} at $U=-1.25$ V.  
 
The electronic isolation of donor-bound states within the depletion region was quantified as illustrated in Figure 1A, by single-electron transport spectroscopy employing the scanning tunneling microscope tip, at 4.2 K. The measured differential conductance $dI/dU$ above donor 1 (Figure 1B, blue squares) exhibited single electron tunneling peaks, from the substrate's impurity band\cite{VanMieghem:1992kx}, to the subsurface donor brought into the bias window by tip-induced band bending, to the tip.  Fitting of the first two peaks to single-electron transport theory\cite{Foxman:1993jq,Mol:2013dj}(Figure 1B, red line) established a coupling $h(\Gamma_{\rm in}+\Gamma_{\rm out}) \ll k_BT \sim 350$ $\mu$eV for donor 1, similar to states in single atom transistors\cite{Lansbergen:2008bs}.  The lowest energy peak ($U\approx -0.8$ V) was found to be bound to donor 1 in spatially resolved $dI/dU$ (Figure 1C).  Higher energy peaks ($U<-1.05$ V) in Figure 1B and 1C are two-electron states (see Supplementary Information). 

Spatially resolved measurements of donor-bound states were carried out using an unconventional scheme.  Constant current imaging of single dopants (see [\onlinecite{Koenraad:2011ed}], references therein, and  [\onlinecite{Sinthiptharakoon:2013il}]) was avoided since it necessarily measures multiple states.  On the other hand, measurement of $dI/dU$ vs. $U$ on a two-dimensional spatial grid would be impractically time consuming for the spatial frequency range $|k_x|,|k_y|\ge 8\pi/a_0$ necessary to avoid frequency aliasing, and desired resolution $\sim 2\pi/50a_0$ to resolve band structure components.  We therefore employed an unconventional two-pass scan to spatially measure individual donor-bound states.  In the first pass of each line, the surface topography $z(x,y)$ was measured at $U=-1.45$ V.  During the second pass of each line, a current $I(\mathbf{r})$ proportional to the probability density $|\Psi(\mathbf{r})|^2$ for a single bound state was measured, at a bias ($U_2=-0.85$ V for donor 1) ensuring one state in the bias window for single-electron transport.  The feedback loop was turned off for the second pass, and the tip position set to $\mathbf{r}=(x,y,z(x,y)-\delta z)$, $\delta z=0.25$ nm closer to the sample relative to the first line (topography).  The value of $\delta z$ was chosen such that $I(U_2)\sim\exp(\kappa z)\sim\Gamma_{\rm out}$, or equivalently, $\Gamma_{\rm out}\ll\Gamma_{\rm in}$.  In this regime, $I(\mathbf{r})\propto|\Psi(\mathbf{r})|^2$, where $\mathbf{r}$ is the central coordinate of the tip apex orbital\cite{Chen:1990cr}.  For donor 1, the measured $z(x,y)$ and $I(\mathbf{r})\propto|\Psi(\mathbf{r})|^2$ are shown in Figure 1D and Figure 1E, respectively.  In Figure 1E, $|\Psi(\mathbf{r})|^2$ vanishes away from the donor, as expected for a measurement of a single bound quantum state.  

\begin{figure*}
\includegraphics{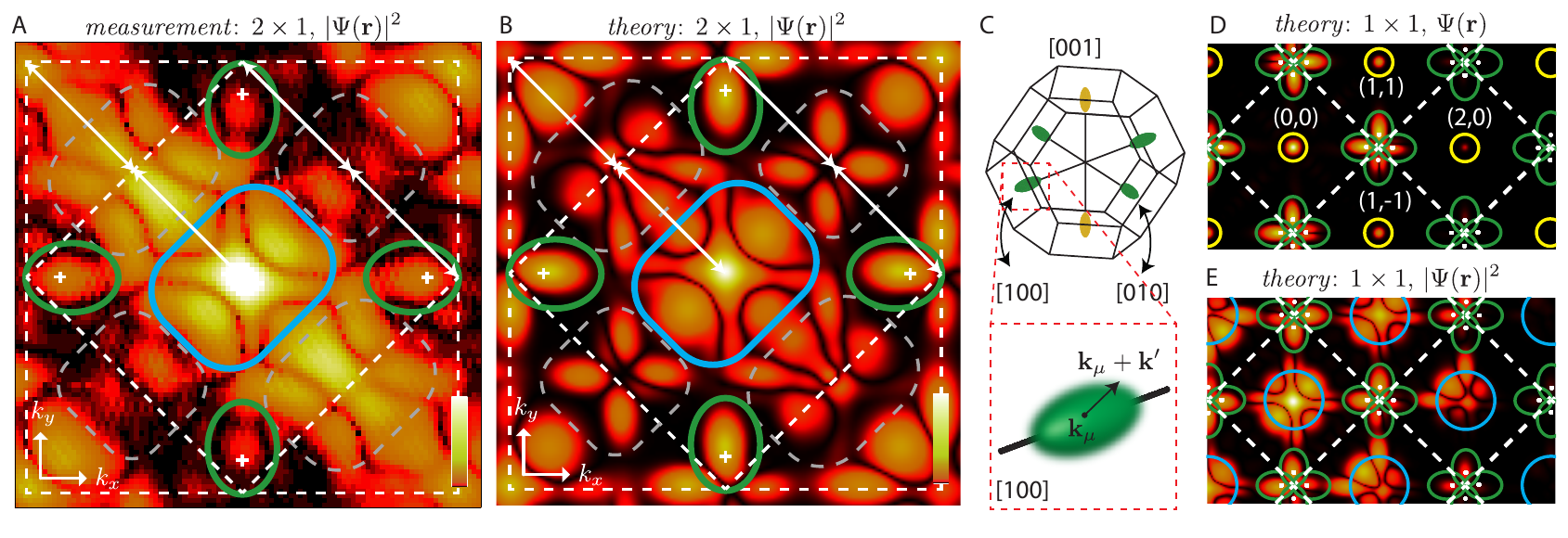}
\caption{\textbf{Valley interference of a single quantum state in reciprocal space: measurement and theory} A.  Fourier amplitudes of measured $I(\mathbf{r})\propto|\Psi(\mathbf{r})|^2$ for donor-bound electron in Figure 1.  Corners of outer dashed square are reciprocal lattice vectors $2\pi/a_0(p,q)$ with $p=\pm 1$ and $q=\pm 1$.  White crosses denote $k_\mu=0.85(2\pi/a_0)(\pm 1,0)$ and $k_\mu=0.85(2\pi/a_0)(0,\pm 1)$.  Ellipsoids structures are found within green boundaries, probability envelope is found within blue boundaries, and $2\times1$ reconstruction-induced features are found within grey dashed boundaries.  B.  Calculated Fourier amplitude of single-electron (D$^0$) ground state probability density $|\Psi(\mathbf{r})|^2$ for $z_0=6.25a_0\approx3.4$ nm deep donor, with a $2\times1$ surface reconstruction.  C.  Distribution of momenta about valley minima in the plane of the surface (green ellipsoids) and perpendicular to it (yellow ellipsoids), and closeup of single valley distribution $F_{x}(\mathbf{k}')$.  D. Overlay of reciprocal space model for D$^0$ ground state wavefunction on atomistic calculation of $\Psi(\mathbf{k})$ using tight-binding method.  Labeled wave vectors have units $2\pi/a_0$.  E. Overlay of reciprocal space model for the D$^0$ ground state probability density on atomistic calculation using tight-binding method.  For D and E, a $1\times1$ reconstruction was assumed for simplicity.  }
\end{figure*}

To investigate valley interference of donor 1, we numerically evaluated the two-dimensional Fourier representation of the measured $I(\mathbf{r})\propto|\Psi(\mathbf{r})|^2$, shown in Figure 2A.  The vertices of the outer dashed box are reciprocal lattice vectors $2\pi/a_0(p, q)$ with $p=\pm 1$, $q=\pm 1$, and $a_0=0.543$ nm.  Figure 2A contains ellipsoid shaped structures, highlighted in green, at positions $\mathbf{k}=k_\mu\hat{x}$ and $\mathbf{k}=k_\mu\hat{y}$ (where $k_\mu= 0.85(2\pi/a_0)$) of silicon's conduction band minima indicated by white crosses.  We term the structure outlined in blue within $k\lesssim 0.5(2\pi/a_0)$ the probability envelope, because it contains the lowest spatial frequencies of the probability density.  Dashed outlined features centered at $2\pi/a_0(\pm1/2, \mp1/2)$ and $2\pi/a_0(\pm1/2, \pm1/2)$ are created by the $2\times 1$ reconstruction, and are related to the probability envelope and ellipsoids, respectively, as evidenced by their $2\pi/a_0(\mp1/2, \pm1/2)$ displacement (white arrows, Figure 2A) relative to them.  A depth $z_0=(5 \pm 1)a_0$ nm ($20 \pm 4$ lattice planes) for donor 1 was empirically obtained by fitting the spatially resolved spectral shift of the conduction band edge to a dielectric screened Coulomb potential\cite{Teichmann:2008bh,Lee:2010ko,Mol:2013dj} (see Supplementary Information).  Shown for comparison in Figure 2B is the Fourier transform of the vacuum tail of the ground state probability density $|\Psi_1(\mathbf{r})|^2$ obtained by sp$^3$d$^5$s$^*$ tight-binding (see Methods).  The predicted ellipsoids and probability envelope features are in very good agreement with measurements in Figure 2A.  

As follows from the discussion below, the ellipsoids and probability envelope features are produced by quantum interference processes associated with quantum superpositions of momenta in six different valleys, in the donor ground state.  Such a state can be written as\cite{Wellard:2005hs} $\Psi_i(\mathbf{r})=\sum_\mu\alpha^i_{\mu}\int{d\mathbf{k}'^3 F_\mu(\mathbf{k}')\phi_{\mathbf{k}_\mu+\mathbf{k}'}(\mathbf{r})}$, where $\boldsymbol{\alpha}_\mu^i$ ($\mu=1...6$) are the valley quantum numbers, and as illustrated in Figure 2C, $F_\mu(\mathbf{k}')$ is the the reciprocal space distribution of the wavefunction over the Bloch functions $\phi_{\mathbf{k}_\mu+\mathbf{k}'}(\mathbf{r})$ about minima $\mathbf{k}_\mu$.  The full Fourier representation of the state is obtained by substituting the representation for the Bloch functions, such that\cite{Wellard:2005hs} $\Psi(\mathbf{r})=\sum_\mathbf{G}\sum_\mu\int d\mathbf{k'}^3\Psi_i(\mathbf{k})\exp(i\mathbf{k}\cdot\mathbf{r})$, where $\Psi_i(\mathbf{k})=\alpha^i_\mu A_{\mathbf{k}-\mathbf{G},\mathbf{G}} F_\mu(\mathbf{k}')$, $\mathbf{k}=\mathbf{G}+\mathbf{k}_\mu+\mathbf{k}'$, and $\mathbf{G}=[p,q,r]=[0,0,0], [1,1,1], [2,0,0], [2,2,0],[3,1,1],...$ (units: $2\pi/a_0$) and their equivalents are the reciprocal lattice vectors with nonzero Bloch amplitudes $A_{\mathbf{k}-\mathbf{G},\mathbf{G}}$ in silicon\cite{Cohen:1966iv}.  

Since $\Psi(\mathbf{k}_\mu)\propto\alpha_\mu F_\mu(0)$, ellipsoid features in $|\Psi(\mathbf{k})|$ at $\mathbf{k}=\mathbf{G}_2\pm \hat{x}k_{\mu}$ and $\mathbf{k}=\mathbf{G}_2\pm\hat{y}k_{\mu}$ (green ellipsoids) are only present when $\alpha^i_{\pm x}\neq 0$ and $\alpha^i_{\pm y}\neq 0$, respectively, while the projected ellipsoid at $\mathbf{G}_2$ (yellow circle) is only present when $\alpha^i_{\pm z}\neq 0$.  For a donor in bulk silicon, the ground state is a spin degenerate valley singlet with $\boldsymbol{\alpha}^1=6^{-1/2}[1,1,1,1,1,1]$.  Hence, the calculated ground state Fourier amplitudes $\Psi_1(\mathbf{k})$ presented in Figure 2D for a depth $6.25a_0$ are in good agreement with the expected reciprocal space representation for a bulk-like donor with valley configuration similar to $\boldsymbol{\alpha}_1$.  The singlet, whose ionization energy is $52$ meV for an arsenic donor, lies 22 meV below a spin-degenerate excited valley triplet with $\boldsymbol{\alpha}^2=2^{-1/2}[1,-1,0,0,0,0]$, $\boldsymbol{\alpha}^3=2^{-1/2}[0,0,1,-1,0,0]$, and $\boldsymbol{\alpha}^4=2^{-1/2}[0,0,0,0,1,-1]$.  The constructive valley interference of the singlet at the arsenic site (the central cell) is responsible for the energy difference.  \cite{Kohn:1955kc,Pantelides:1974ew}

\begin{figure}
\includegraphics{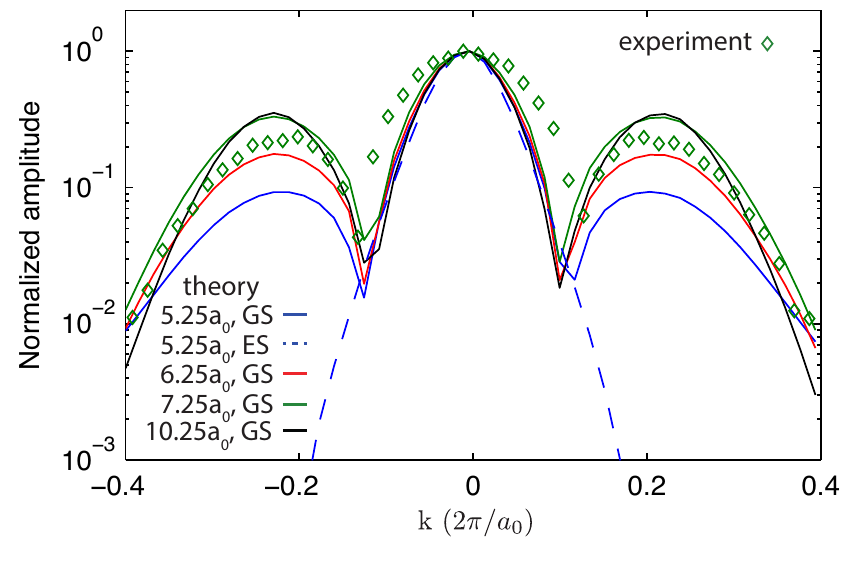}
\caption{\textbf{Valley population} Comparison of measured (green diamonds) and theoretical (solid lines) reciprocal space profile of $|\Psi_1(\mathbf{r})|^2$ for donor ground state, along 110 direction in reciprocal space.  The height of the central peak ($k=0$) relative to side peaks ($k=0.15(2\pi/a_0)(\pm 1,\pm 1)$) increases with decreasing depth from $10.25a_0$ (black line), $7.25a_0$ (green line), $6.25a_0$ (red line), to $5.25a_0$ (blue line), indicating repopulation from $x$ and $y$ valleys into $z$ valleys with decreasing depth.  The calculated first excited state, shown for $5.25a_0$ (blue dashed line), populates only the $z$ valley, and hence, has no side lobes.}
\end{figure}

\begin{figure*}
\includegraphics{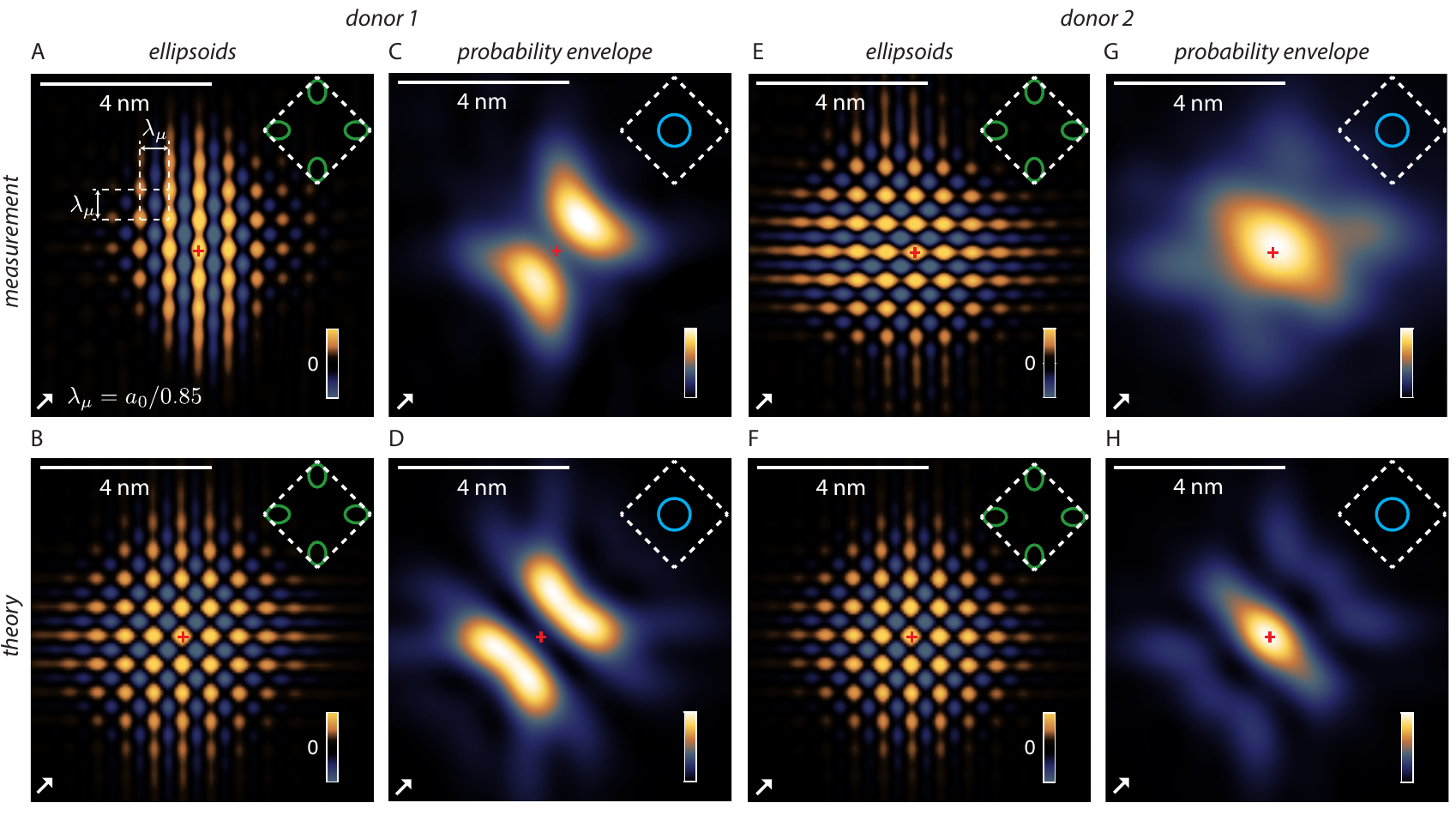}
\caption{\textbf{Valley interference of a single quantum state in real space: measurement and atomistic theory} A.  Measured real-space representation of four ellipsoid features at $k\approx k_\mu$ for donor 1 in Figure 2, obtained by frequency selective inverse Fourier transform of Figure 2A.  Scale:(-15,15) pA.  Arrow denotes [110] dimer direction.  Red cross denotes ion position determined from real-space ellipsoid pattern, as discussed in Supplementary Information.  B.  Predicted real-space representation of four ellipsoid features based on atomistic calculations.  The best match was obtained for $z_0=6.25a_0\approx3.4$ nm deep donor.  C. Measured real-space representation of probability envelope.  D.  Predicted real-space representation of type-A probability envelope.  E.  Same as A, but for measured donor 2 with type-B probability envelope.  F. Predicted real-space representation of four ellipsoid features based on atomistic calculations.  The best match was obtained for a $z_0=6.75a_0\approx3.7$ nm deep donor.  G. Same as B, but for type-B probability envelope.  H.  Predicted real-space representation of probability envelope based on atomistic calculation for F.   Inset: Corresponding region in first Brillouin zone (dashed lines).}
\end{figure*}

Two quantum interference processes are of particular interest in the Fourier representation of Figure 2A and 2B, which maps components $\exp(-i\mathbf{k}'\cdot\mathbf{r})$ in $\Psi_i^*(\mathbf{r})$ and $\exp(i\mathbf{k}''\cdot\mathbf{r})$ in $\Psi_i(\mathbf{r})$ to $\exp(i(\mathbf{k}''-\mathbf{k}')\cdot\mathbf{r})$ in $|\Psi_i(\mathbf{r})|^2=\Psi_i^*(\mathbf{r})\Psi_i(\mathbf{r})$.  First, ellipsoids at $\mathbf{k}\approx\pm\hat{x}k_\mu$ ($\mathbf{k}\approx\hat{y}k_\mu$) in $|\Psi_i(\mathbf{r})|^2$, overlaying atomistic calculations in Figure 2E, arise from cross-product terms of $z$ valleys and $x$ ($y$) valleys.  Consequently, the ellipsoids in the Fourier decomposition of $|\Psi(\mathbf{r})|^2$ are direct evidence for the presence of all three valleys in the measured (Figure 2A) and atomistically calculated (Figure 2B) subsurface donor ground state.  From the same mapping, it follows that $x$, $y$, and $z$ valleys contribute to the peak at $k=0$, while only $x$ and $y$ valleys contribute to the side peak at $k=0.15(2\pi/a_0)(\pm 1,\pm 1)$. 

The relative heights of the central and side peaks in the measured and calculated probability envelopes were compared to characterize the deviation of donor 1 from the $\boldsymbol{\alpha}^1$ (bulk) valley configuration.  The electric field in experiments, estimated in the Supplementary Information to be $\mathcal{E}_z=-0.3\pm1.9$ MV/m using tip-height dependent $dI/dU$ measurements, is low enough to be safely neglected in calculations of donor-bound states\cite{Rahman:2011ev}. Calculations for decreasing depths $10.25a_0$ to $5.25a_0$ presented in Figure 3 demonstrate an increasing contribution of the central peak signaling an increase in the $z$-valley population relative to $x$ and $y$ valleys.  This arises from the competition between the interface, which introduces a valley-orbit potential\cite{Goswami:2006bb,Saraiva:2009gh,Yang:2013if} and depopulates the $x$ and $y$ valleys with lighter masses parallel to the interface, and the donor ion's valley-orbit potential, which equalizes all six valleys.  The side peak amplitude saturates at $7.25a_0$ indicating that the valleys approach the bulk configuration.  Shown in Figure 3 (green diamonds), the measured donor's profile agrees well with calculations for $6.25a_0$ at zero field, in excellent agreement with the empirically determined depth $z_0=(5\pm 1)a_0$ of donor 1.  We estimate that the $z$-valley configuration of the measured donor differs by only $\sim 5 \%$ compared to a donor in bulk silicon ($33.3 \%$), since states calculated for $7.25a_0$ and $6.25a_0$ depths have $z$-valley populations of $36 \%$ and $40 \%$, respectively (See Supplementary Information).  The wider central peak in measurements could be the result of enhanced lateral localization due to image charges associated with dielectric mismatch for the 3 nm deep donor, not taken into account in the theory.  Overall, the agreement of the reciprocal space profile with the theoretical description from the calculations is remarkable.  Note that low-pass response of the STM tip orbital is not expected for such low spatial frequencies $k \lesssim 0.21 (2\pi/a_0)$ ($\lambda \gtrsim 2.6$ nm).  

The real-space representation of the ellipsoids, shown in Figure 4A and obtained by digital Fourier filtering, resembles a low frequency s-like envelope modulated by x and y-directed standing wave patterns with wavelength $\lambda_\mu \approx 2\pi/k_\mu = 0.65$ nm.  As described above, the ellipsoids in $|\Psi_1(\mathbf{r})|^2$ arise from cross terms between the $x$ ($y$) valleys, which oscillate at $\mathbf{k}=k_\mu\hat{x}$ ($\mathbf{k}=k_\mu\hat{y}$), and the $z$ valley.  Consequently, their real-space representation can be easily shown (see Supplementary Information) to have the form $F_z(\mathbf{r})(F_x(\mathbf{r})\cos(k_\mu x)+F_y(\mathbf{r})\cos(k_\mu y))$, where $F_i(\mathbf{r})=\int{d^3\mathbf{k}'F_i(\mathbf{k}')\exp(i\mathbf{k}'\cdot\mathbf{r})}$ is the donor ground state envelope function.  The maximum of the real-space oscillation pattern, labelled with a red cross in Figure 4, identifies the position of the ion.  Notably, the appearance of elongation along [100] and [010] directions (Figure 4A) is also predicted by atomistic theory (Figure 4B). The oscillations observed in Figure 4A/4B are associated with superpositions of momenta, not unlike Friedel oscillations of metallic surface states observed by STM at impurities and step edges\cite{CROMMIE:1993vw,Sprunger:1997ck}.  The present case differs because the superpositions occupy six valleys along three orthogonal [001] directions, rather than a two-dimensional Fermi surface.

Like the ellipsoids, the probability envelope feature introduced in the discussion of Figure 2 arises from valley interference.  Therefore, it is not surprising that the real-space representation of the probability envelope in Figure 4C has an unusual shape characterized by a node along $x=-y$.  This node, predicted by calculations (Figure 4D), contrasts the simple envelope of oscillations in Figure 4A/4B.  Notably, a survey of fifty subsurface donors with indistinguishable spectral signatures revealed 24 ``type-A'' envelopes typified by donor 1 in Figure 4C/4D, and 26 ``type-B'' envelopes typified by donor 2 in Figure 4E/4G.  While the ellipsoid interference pattern of the type-B donor (Figure 4E) resembles that of type-A donor (Figure 4A), the probability envelope of the type-B donor (Figure 4G) has a protrusion along $x=-y$.  The depth of donor 2 was not measured due to a technical problem.

The type-A and type-B probability envelopes are at first peculiar.  An examination of tight-binding calculations of $|\Psi(\mathbf{r})|^2$ for donors occupying successively deeper planes revealed a sequence A, A, B, B, A, A, B, B for donor depths $5.00a_0$, $5.25a_0$, $5.50a_0$, $5.75a_0$, $6.00a_0$, $6.25a_0$, $6.50a_0$, and $6.75a_0$.  Both the calculated ellipsoids (Figure 4F) and probability envelope (Figure 4H) of the $6.75a_0$ type-B donor are in excellent agreement with measurements in Figure 4E and Figure 4G, respectively.  Moreover, the calculated probability envelope for a fixed donor depth was found to depend sensitively on the lattice plane below the surface where it was evaluated.  The two different probability envelopes are therefore a manifestation of (1) rapid spatial variation of $|\Psi(\mathbf{r})|^2$ along $z$ with lattice and valley spatial frequencies (and their harmonics), and (2) the relative sensitivity of the tip to the topmost atomic planes.  Nevertheless, the spatial average of type-A and type-B probability envelopes is s-like, as expected from effective mass\cite{Kohn:1955kc}.  We found an uncertainty $\sim a_0$ in the depth determination, considerably larger than the uncertainty $<0.25a_0$ required to assign a donor to a crystal lattice plane.  Nevertheless, the empirically observed 48$\%$ / 52$\%$ distribution is consistent with theory, assuming (as expected) a random distribution of donor depths.

Any interface-induced mixing with excited valley-orbit states enhances the spin-lattice relaxation rate\cite{Roth:1960co,Hasegawa:1960ey}. The nearly bulk-like valley interference pattern observed for $\sim 3$ nm deep donors is a direct indication of small mixing with valley-orbit excited states.  We therefore expect that arsenic donors whose depth exceeds 3 nm (approximately 1 effective Bohr radius) would have spin-lattice relaxation rates $T_1^{-1}$ only slightly larger than bulk donors.  For magnetic fields of practical interest\cite{Morello:2010ga}, $T_1$ of subsurface donors will still greatly exceed the spin coherence time of electrons in isotope purified silicon\cite{Tyryshkin:2011fi}.  However, proximity to non-ideal interfaces could introduce other relaxation or decoherence processes.  Moreover, the observed interference in Figure 4 is precisely the phenomena predicted to render two-qubit gates sensitive to atomic-scale variations in donor position\cite{Koiller:2001gw,Wellard:2005hs}.  This can be thought of as a direct consequence of phase mismatch in interfering terms in the exchange interaction $J(\mathbf{R}) = \int \Psi^*(\mathbf{r}_1-\mathbf{R})\Psi^*(\mathbf{r}_2)V_{ee}(|\mathbf{r}_1-\mathbf{r}_2|)\Psi(\mathbf{r}_1)\Psi(\mathbf{r}_2-\mathbf{R})$ between donors separated by a displacement $\mathbf{R}$, each having lattice-incommensurate ($\lambda = 2\pi/k_\mu = 0.65$ nm) spatial oscillations.  

The spatially resolved single electron transport demonstrated herein provides a new level of access to valley physics.   We have observed electronic valley quantum interference for a single donor atom whose depth and electric field have been independently empirically determined, capabilities not available in single-electron transport spectroscopy of donors in nanoscale transistors\cite{Lansbergen:2008bs,Roche:2012gb,Fuechsle:2012bl}.  We have identified that a donor $2.85\pm0.45$ nm from an interface and in a low field $\mathcal{E}_z=0.3\pm1.9$ MV/m has an electronic valley population deviating by only $\sim 5$\% compared with a donor in bulk silicon.  This new level of understanding is essential to the engineering of addressable single-qubit two-qubit operations in silicon donor devices\cite{Calderon:2006ha,Rahman:2007ev,Rahman:2009kj,Koiller:2001gw,Wellard:2005hs}, in order to exploit silicon's remarkable coherence\cite{Tyryshkin:2011fi,Steger:2012ev}.

\section*{Acknowledgement}

The authors would like to thank J. Verduijn for helpful discussions.  This work is supported by the European Commission Future and Emerging Technologies Proactive Project MULTI (317707) and the ARC Centre of Excellence for Quantum Computation and Communication Technology (CE110001027), and in part by the U.S. Army Research Office (W911NF-08-1-0527).  This work is part of the research program of the Foundation for Fundamental Research on Matter (FOM), which is part of the Netherlands Organization for Scientific Research (NWO).  S.R. acknowledges a Future Fellowship (FT100100589). M.Y.S acknowledges a Federation Fellowship.  

\section*{Author contributions}
J.S., J.A.M. and S.R. designed and conducted the experiments. R.R. performed the multi-million atom calculations. J.S., L.C.L.H. and S.R. made the key contributions to the Fourier analysis.  All the authors contributed to analysis and writing the paper.

\appendix

\section*{Methods}

Samples where single-electron tunneling through donor-bound electronic states was observed were prepared by flash annealing a commercial n-type (arsenic doped) silicon wafer with resistivity $0.004-0.001$ ohm-cm to a temperature $\sim 1050$ $^\circ$C for 10 seconds, a total of 3 times.  After the final flash anneal, the temperature was rapidly quenched to 800 $^\circ$C, followed by slow (1 $^\circ$C/s) cooling to 340 $^\circ$C, producing a $2\times1$ surface reconstruction.  Hydrogen passivation was carried out by dosing with 9 monolayers of atomic hydrogen.  This flash anneal procedure is known from secondary ion mass spectroscopy to deplete the upper $\sim 10$ nm of the wafer of arsenic dopants\cite{Pitters:2012gt}, but is shallow enough to maintain sufficient coupling to the conduction impurity band to obtain a measurable single-electron tunneling current through donor-bound states.  Donors found by subsurface dopant imaging, as described in the main text, were residual impurities statistically distributed throughout the depletion region\cite{Pitters:2012gt,Sinthiptharakoon:2013il}.  No implantation was performed.  No donors were found for samples flashed 3 times at 1200 $^\circ$C, which are expected to have too deep ($\sim 100$ nm) a surface depletion and too few residual arsenic dopants\cite{Pitters:2012gt}. 

Measurements were performed using an Omicron low temperature scanning tunneling microscope (LT-STM) operating in ultra-high vacuum at a temperature of 4.2 K.  Current $I$ was measured as a function of sample voltage $U$ using ultra-low noise electronics, and $dI/dU$ was obtained by numerical differentiation.  Spatially resolved measurements of donors were obtained on frames 20 nm $\times$ 20 nm in size, containing a single subsurface dopant, with a spatial resolution of 0.02 - 0.04 nm.  Fine calibration of reciprocal lattice vector positions was carried out by Fourier transforming topographies (see Supplementary Information) acquired simultaneously with quantum states, using the multi-line scan technique.  

\section*{Supplementary Information}
\subsection{Conventional subsurface dopant imaging}

Conventional constant current subsurface dopant imaging\cite{Koenraad:2011ed,Marczinowski:2007dt} at a bias $U=-1.25$ V was used to establish that subsurface donors were present in the fabricated sample.  The signature of bound states is the localized protrusion seen in Figure A.1.  As apparent from Figure 1B in the main text, localized states at $U=-0.78$ V, $U=-1.05$ V and $U=-1.09$ V all contribute to the constant current image obtained at $U=-1.25$ V.   Moreover, the continuum into which above band-edge tunneling necessarily occurs in constant current imaging is typically disturbed by the presence of impurities\cite{Marczinowski:2007dt}.  

The combination of tunneling from multiple bound states and the locally disturbed continuum, makes constant current imaging at $U=-1.25$ V shown in Figure A.1 unsuitable for probing single quantum states of subsurface donors.  High resolution measurement of discrete single states was accomplished by the two-pass scheme discussed in the main text.  Our two-pass scheme employs topography taken at a sample bias $U=-1.45$ V (Figure 1D, main text), which shows scant evidence for the subsurface donor.  Donors 1 and 2 discussed in the main text were found employing large-area (75 nm $\times$ 75 nm) scans and the two-pass technique.  

\begin{figure}
\includegraphics{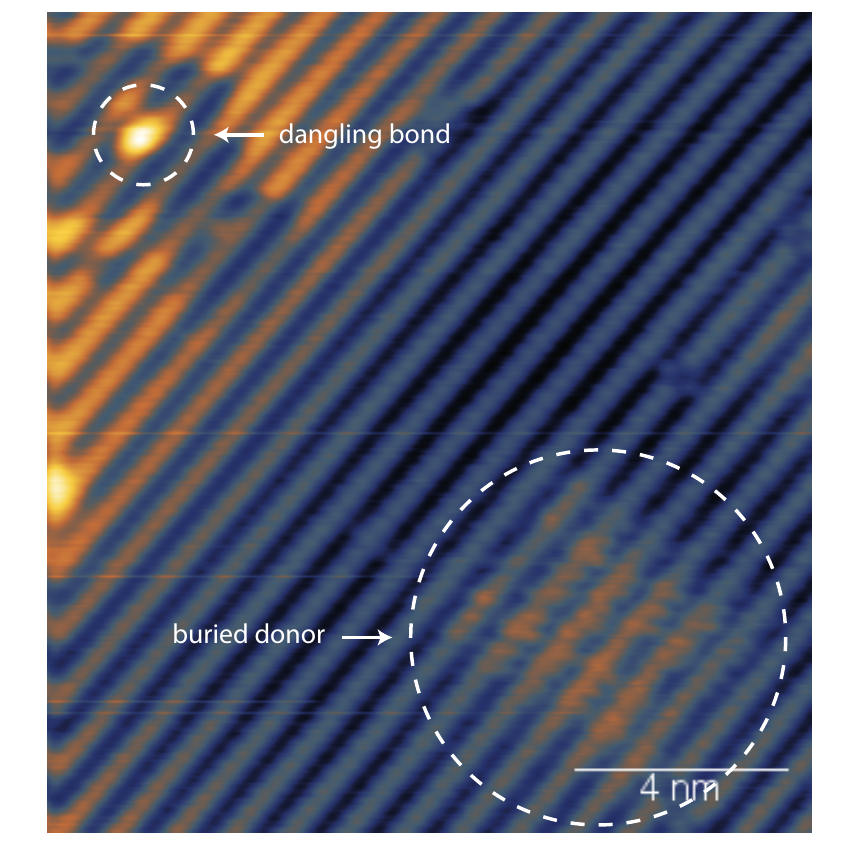}
\begin{flushleft}
{Fig. A.1. Constant current image at $U=-1.25$ V and 4.2 K.  The signature of a buried donor is circled in the bottom right corner. }
\end{flushleft}
\end{figure}

\subsection{Donor Depth}

The depth $z_0$ of donor 1 was estimated using a standard method\cite{Lee:2010ko, Teichmann:2008bh, Mol:2013dj}, whereby the Coulomb potential produced in the donor's ionized state ($U>-0.75$ V) was mapped spatially at the semiconductor surface, and fit to an analytic expression that depends on the dopant depth.  Following previous work\cite{Teichmann:2008bh, Mol:2013dj}, we measured the spatial variation of a band edge in the donor's ionized state.  Away from the donor (Figure A.2.A), $I$ vs. $U$ measurements show that the conduction band edge appears from the noise floor at $U\approx 0.20$ V.  For a tip position directly above the donor atom $I$ vs. $U$ measurements show that the sample voltage at the onset of tunneling is shifted downwards towards $U\approx 0.14$ V, by approximately $60$ mV.  

The sample voltage $U_C$ required to obtain a tunneling current 0.1 pA was mapped as a function of position, and is shown in Figure A.2.B.  Directly at the tip position above the donor, the smallest voltage $U_C\approx0.14$ V is obtained, while away from the donor, the conduction band edge is located at $U_C \approx 0.2$ V.  Because $U_C\approx0.1 - 0.2$ V exceeds the flatband voltage $U_{FB}=-0.83\pm 0.034$ V (see section 4), the surface is depleted due to tip-induced band bending during the conduction band edge measurement.  It is therefore appropriate to fit the spatial variation to a dielectric-screened Coulomb potential\cite{Lee:2010ko, Teichmann:2008bh, Mol:2013dj}, ignoring screening by  free carriers.  

The potential energy at a position $\mathbf{r}=(x,y,z)$ due to an ionized dopant at $\mathbf{r}_0= (x_0,y_0,z_0)$ inside a semiconductor occupying the half-space $z\leq 0$ is obtained from the method of image charges\cite{Hao:2009eq}.  Inside the semiconductor, 
\begin{equation}
V(\mathbf{r}) = \frac{-e^2}{4\pi\epsilon_0\epsilon_{2}}  \bigg[\frac{1}{|\mathbf{r}-\mathbf{r}_0|} - \frac{\epsilon_1-\epsilon_2}{\epsilon_1+\epsilon_2}\frac{1}{|\mathbf{r}-\mathbf{r}_1|}\bigg]
\end{equation}
is obtained, where $\epsilon_2$ and $\epsilon_1$ are the relative dielectric permittivity of the semiconductor and environment, respectively, and $\mathbf{r}_1 = (x_0,y_0,-z_0)$ is the position of the image charge.  Evaluating the Coulomb potential shift to the conduction band edge at the interface ($z=0$), the spectral shift to the conduction band

\begin{equation}
\Delta U_C(x-x_0,y-y_0)=\frac{-e/4\pi\epsilon_0\epsilon_{\textrm{eff}}}{\sqrt{x^2+y^2+z_0^2}},
\end{equation}
is obtained, where $\epsilon_{\textrm{eff}}=(1+\epsilon_{2})/2$ is an effective dielectric constant for a semiconductor-vacuum interface.  This expression was fit to the measured $U_C(x,y)$ for donor 1 in Fig A.2.B.  Profiles for both the measured and fit spectral shift are presented in Figures A.2.C and A.2.D for perpendicular ($x$ and $y$) directions, showing excellent agreement between the model and the experiments.  For donor 1, $|z_0|=2.85\pm 0.45\textrm{ nm }\approx (5\pm 1)a_0$ and $\epsilon_{\textrm{eff}}=6.1\pm 1$ were obtained.  The effective dielectric constant closely matches the value $\epsilon_{\textrm{eff}}=(1+11.7)/2 = 6.35$ expected for a silicon/vacuum interface.  A technical problem (STM tip crash) prevented us from measuring the band edge shift for donor 2 discussed in the main text.

\begin{figure}[h]
\includegraphics{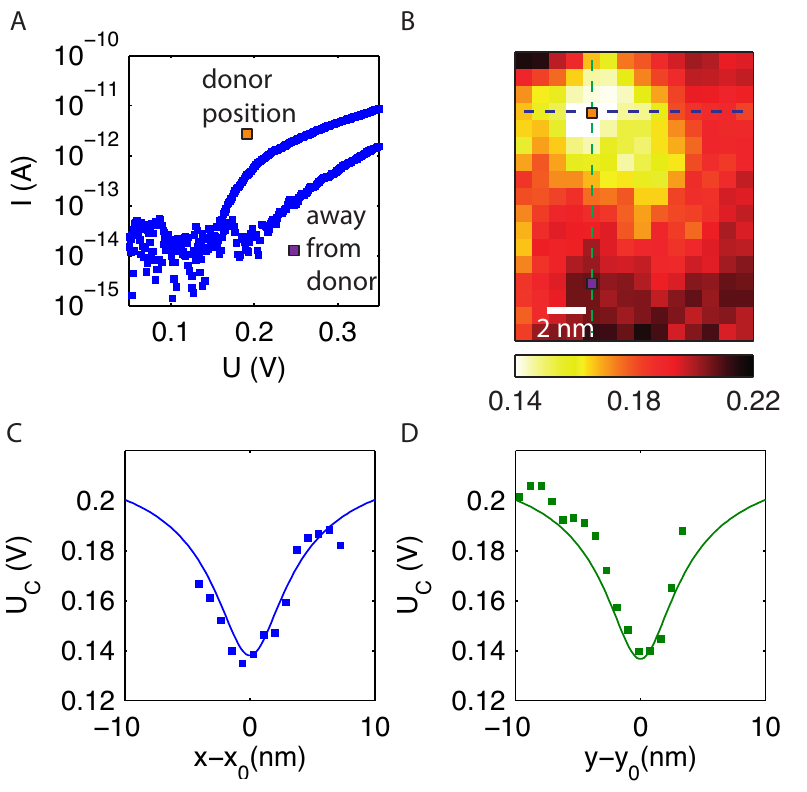}
\begin{flushleft}
{Fig. A.2.A. $I$ vs. $U$ measurements above the donor and away from the donor showing the onset of tunneling into the conduction band edge at $U \approx 0.14$ V and $U \approx 0.20$ V, respectively. B.  Map of measured sample voltage $U_C$ at the onset of tunneling into the conduction band.  C. Profile of measured (blue squares) and fit (blue solid line) voltage $U_C$ at the onset of tunneling into the conduction band, along $x$ direction.  D.  Profile of measured (green squares) and fit (green solid line) voltage $U_C$ at the onset of tunneling into the conduction band, along $y$ direction.   }
\end{flushleft}
\end{figure}

\subsection{Single-electron transport spectroscopy}

In this section, the single-electron tunneling spectra measured for tip positions above donors 1 and 2 are compared to the theory of single-electron transport in the weak coupling regime\cite{Foxman:1993jq}, which describes similar experiments on boron acceptors in silicon\cite{Mol:2013dj}, and transport experiments on donors in single-atom transistors\cite{Lansbergen:2011eua}.  

\begin{figure*}[ht]
\includegraphics{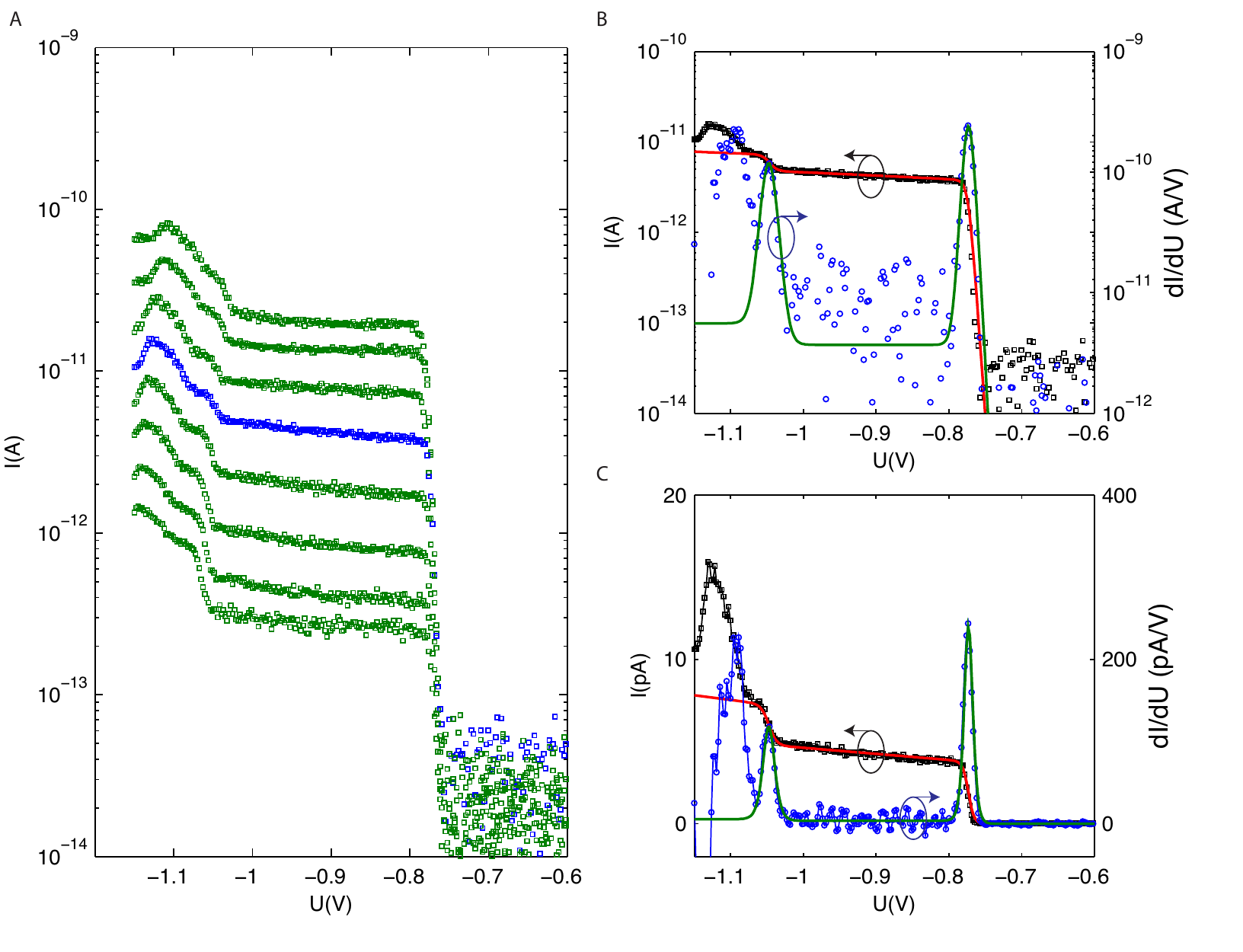}
\begin{flushleft}
{Fig. A.3.A.  Measured single electron tunneling current for different tip heights successively 28 pm higher above donor 1 in the main text.  B.  Measured current $I$ (black squares), numerical derivative $dI/dU$ (blue circles), least-square fit to current (red line), and corresponding derivative of fit (green line). C. Same as B but plotted on a linear scale.  }
\end{flushleft}
\end{figure*}

The $I$-$U$ characteristic for tip position above donor 1 is presented for several tip heights in Figure A.3.A.  Sequentially from the top to bottom, data correspond to 8 different tip heights each successively $28$ pm higher above the dopant.  Blue data points are for the tip height presented in Figure 1B in the main text.  The nearly fixed offset (on a logarithmic scale) between curves obtained at different tip heights reflects the exponential decay of the wavefunction into vacuum\cite{Tersoff:1985bw}, which can only be observed in the single-electron tunneling regime described by $\Gamma_{\rm out} \ll \Gamma_{\rm in}$ (see Figure 1B).  From $\Gamma_{\rm out} \ll \Gamma_{\rm in}$ we have $I(\mathbf{r})\propto|\Psi(\mathbf{r})|^2$.  

On a logarithmic scale, the linear drop of current into the noise floor ($\sim 50$ fA) is the clear experimental signature of thermal broadened single-electron transport\cite{Foxman:1993jq}.  State energies $E_i=e\alpha U_i$ were obtained by least-squares fitting of the lever arm $\alpha$ and voltage $U_i$ to the well-known line shape for single-electron tunneling\cite{Foxman:1993jq}, as recently employed to describe similar experiments on Boron acceptors in silicon\cite{Mol:2013dj}.  The measured current $I(U)$ was fit to $I=\int_{0}^{U}(\partial I(U')/\partial U') dU'$, where
\begin{equation}
\frac{\partial I}{\partial U} = \sum_{i}A_i\int_{-\infty}^{\infty}{dE\frac{\frac{1}{2}h\Gamma_i\times \cosh^{-2}(E/2k_BT)}{(\frac{1}{2}h\Gamma_i)^2+(\alpha e (U-U_i)-E)^2}},
\end{equation}
and $\Gamma_i$ and $A_i$ are lifetimes and amplitudes for states $i=1,2$.  The proportionality constant $A_i$ was allowed to vary linearly with tip height to account for a small positive differential conductance that arises either by field-induced lowering of the barrier to the STP tip with decreasing bias $U$.  Results for the fit of donor 1 are given in Table I.   

\begin{table}
\begin{center}
\begin{tabular} {| c | c | c |}
\hline
donor & donor 1 & donor 2\\
\hline
\hline
$U_1$ (V) & $-0.7742 \pm 0.0002$ & $-0.7567 \pm 0.00051$ \\
$U_2$ (V) & $-1.0488 \pm 0.0005$ & $-1.0241 \pm 0.00043$ \\
$\alpha_1$ & $0.0878 \pm 0.0033$ & $0.0632 \pm 0.0110$ \\
$\alpha_2$ & $0.0627 \pm 0.0049$ & $0.0813 \pm 0.0110$ \\
$h\Gamma_1/k_BT$ & - & $5.1264\pm 1.0433$ \\
$h\Gamma_2/k_BT$ & - & $2.8345\pm 0.6472$ \\
$\frac{1}{2}h\Gamma_1/3.5k_BT$ & - & $0.7323\pm 0.1490$ \\
$\frac{1}{2}h\Gamma_2/3.5k_BT$ & - & $0.4049\pm 0.0925$ \\
$E_2-E_1$ (meV) & $21\pm 2$ & $20 \pm 3$ \\
\hline
\end{tabular}
\begin{flushleft}
{Table I. Results for least-squares fits of data in Figure A.3 and A.4 to model discussed in methods. Donor 1 was fit to purely thermally broadened resonances.  }
\end{flushleft}
\end{center}
\end{table}

Figure A.3.B shows the least-square fits of the measured current $I$ (black squares) to the model discussed above (red line), as well as numerically differentiated conductance $dI/dU$ (blue circles), and the derivative of the least square fit (green line). The same results plotted on a linear scale are presented in Figure A.3.C.  

\begin{figure*}[ht]
\includegraphics{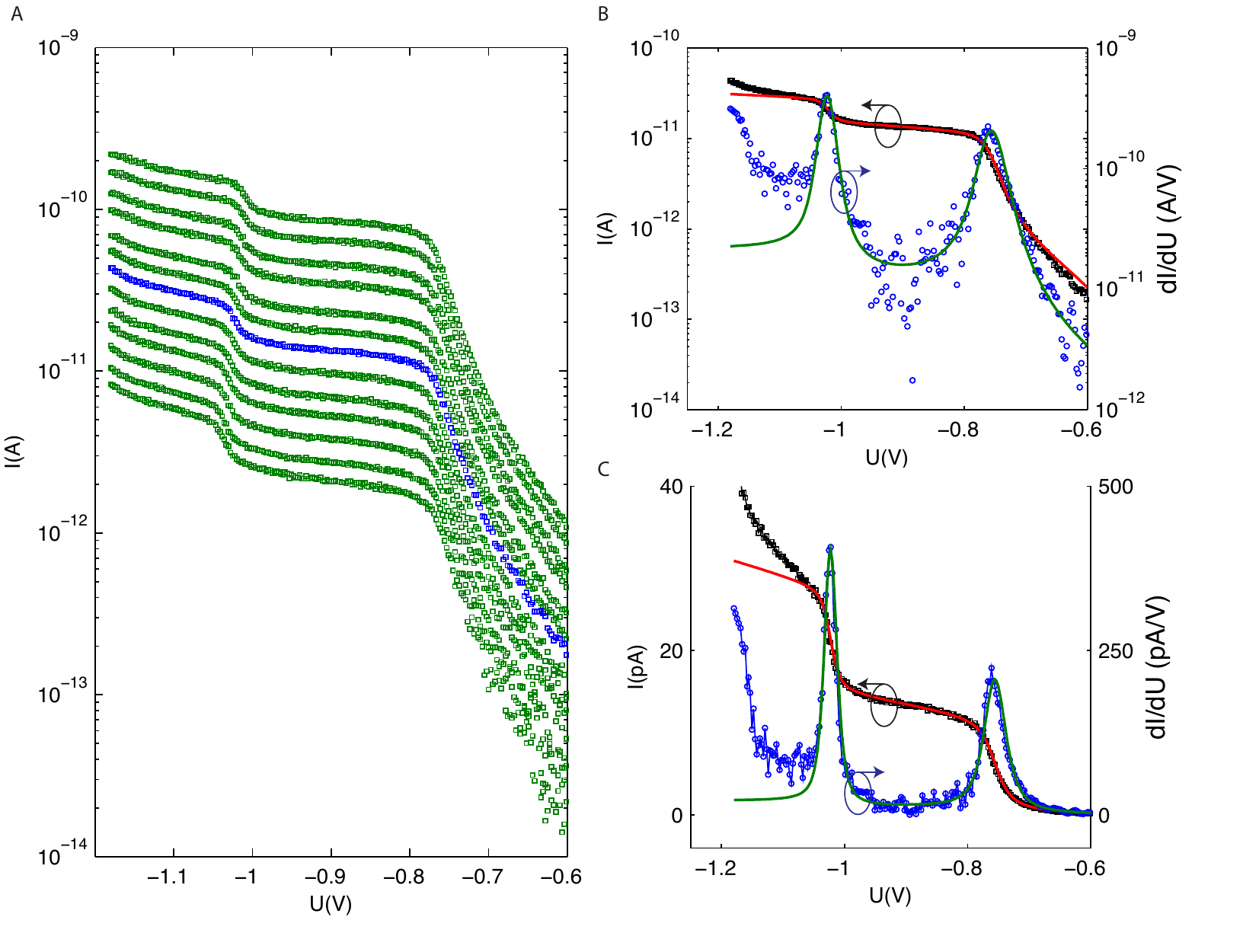}
\begin{flushleft}
{Fig. A.4.A.  Measured single electron tunneling current for different tip heights successively 14 pm higher above donor 2 in the main text.  B.  Measured current $I$ (black squares), numerical derivative $dI/dU$ (blue circles), least-square fit to current (red line), and corresponding derivative of fit (green line). C. Same as B but plotted on a linear scale.  }
\end{flushleft}
\end{figure*}

The single-electron tunnel current spectrum of donor 2 from the main text is presented for several tip heights in Figure A.4.A.  Data (green squares) are shown for 13 tip heights each successively $14$ pm higher above the dopant.  The upturn in the logarithm of the tunnel current indicates that the single-electron transport has an additional lifetime component contributing to its lineshape\cite{Foxman:1993jq}.   

The curve plotted in blue squares in Figure A.4.A was fit to a model of thermally broadened Lorentzians (red line) in the sample voltage range -0.65 V to -1.08 V. Figure A.4.B shows the least-square fits of the measured current $I$ (black squares) to the model (red line), as well as numerically differentiated conductance $dI/dU$ (blue circles), and the derivative of the least square fit (green line). The same results plotted on a linear scale are presented in Figure A.4.C.  Results for the fit of donor 2 are given in Table I.  Since the lifetime broadening $\Gamma$ of donor 2 does not increase with decreasing tip height, its dominant coupling is to the conduction impurity band.  The larger coupling of donor 2 compared with donor 1 could arise if donor 2 is deeper beneath the silicon surface relative to donor 1, or because of a decrease in the depth of the annealing-induced depletion in the vicinity of donor 2 compared with donor 1.  As mentioned in section II, the depth of donor 2 was not extracted due to technical difficulties (tip crash), so it is not possible to comment further on the origin of the stronger coupling of donor 2.

Two $dI/dU$ peaks for $U\lesssim -1.0$ V in Figure A.3.B and A.4.B were consistently found in the spectra of subsurface donors.  For donor 1 in Figure 1C of the main text (depth $2.8\pm 0.45$ nm), the states merge into tip-induced quantum dot states away from the donor, and can therefore directly be identified as two-electron states above the donor.  For donor 1, the lowest energy two electron state ($U_2=-1.051$ V) is a singlet, while the state at $U_3=-1.094$ V is the two-electron triplet likely involving an orbital excited state with a different valley configuration.  The absolute energy difference $E_3-E_2=e\alpha(U_3-U_2)$ was estimated using $\alpha = 0.065\pm 0.005$ obtained at $U=U_2$.  We obtain $E_3-E_2=2.8\pm 0.2$ meV, slightly higher than for donors near a silicon-SiO$_2$ interface in a nanoscale transistor\cite{Lansbergen:2011eua}. 

No strong single-electron (D$^0$) excited state $dI/dU$ peaks were observed for donor 1 or donor 2. Since $\Gamma_{\rm out}\ll\Gamma_{\rm in}$ in our experiment, single-electron excited states would only observed if they have both (1) appreciably different couplings $\Gamma_{\rm out}$ to the tip and (2) inelastic relaxation rates less than $\Gamma_{\rm out}$, conditions that may not be met for D$^0$ valley excited states of donors with bulk-like valley configurations.

\subsection{Electric field experienced by donor}

The electric field $\mathcal{E}_z$ perpendicular to the interface during measurement of the $D^0$ state was evaluated directly from the measured variation $dU_i/dz$ in peak voltage $U_i$ with tip height $z$ in Figure A.3.A.  Simple electrostatic arguments for tip-induced band bending were employed\cite{Feenstra:1987ic}, identical to those describing metal-oxide-semiconductor capacitors.  We find that the D$^0$ state is measured in a regime $\mathcal{E}_z = (-0.3 \pm 1.9)$ MV/m, well below the field $\sim20$ MV/m theoretically predicted to contribute to coherent valley de-population for an arsenic donor $\sim 3$ nm from an interface\cite{Rahman:2011cu}.  These electrostatic arguments were independently validated by extracting the vacuum tunneling barrier energy\cite{Loth:2007hw,Garleff:2011dl,Mol:2013dj} from measured variation $dI/dz$ in the tunnel current $I$ with tip height $z$ in Figure A.3.A.

\begin{figure}[h]
\includegraphics{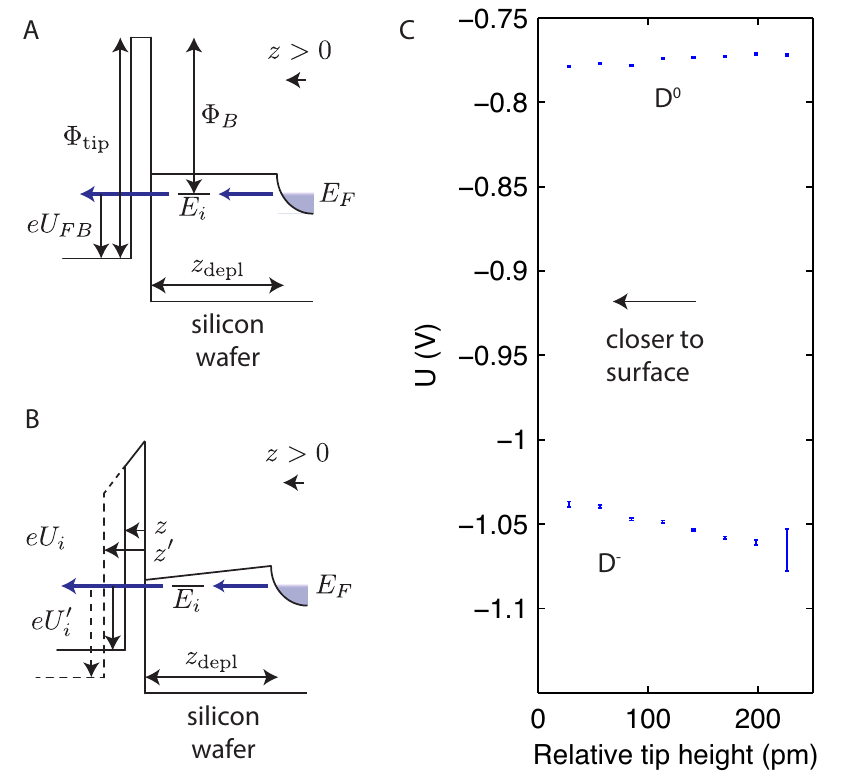}
\begin{flushleft}
{Fig. A.5.A.  Illustration of flatband condition for zero electric field in the sample when the sample voltage $U=U_{\rm FB}$ exactly compensates the difference in work function between the tip and sample.   B.  For states of energy $E_i$ on resonance with the $E_F$ for $U_i < U_{\rm FB}$, the derivative $dU_i/dz$ determines the vacuum electric field.  C.  Extracted dependence of D$^0$ peak voltage $U_1$ and D$^-$ peak voltage $U_2$ for donor 1.  }
\end{flushleft}
\end{figure}

Electron tunneling from a localized state in a semiconductor to the tip is illustrated in Figure A.5.A for the flatband condition.  In this condition, the applied sample bias $U=U_{FB}$ exactly cancels the difference between the tip work function $\Phi_{\rm tip}$ and the sample Fermi energy $E_F$, the electric field $\mathcal{E}_z$ in the semiconductor is zero, and the vacuum tunneling barrier is rectangular.  At the flatband condition, a localized state (energy $E_i$) on resonance with $E_F$ produces a peak in $dI/dU$ at a voltage $U_i=U_{FB}$ that is independent of tip height $z$.  The situation is different away from flatband, when $\mathcal{E}_z\neq 0$.  As illustrated for $U_i<U_{FB}$ in Figure A.5.B, a trapezoidal vacuum barrier and electric field $\mathcal{E}_z>0$ attracting electrons to the surface are obtained.  From simple electrostatics, a peak found at a bias $U_i<U_{\rm FB}$ for a tip height $z$ shifts in voltage to $U_i'=U_i-\mathcal{E}_{z,\rm{vac}}(z'-z)$ at a different tip height $z'$, where $\mathcal{E}_{z,\rm{vac}}=-dU/dz$ is the vacuum electric field.  This is geometrically illustrated in Figure A.5.B for $z'>z$.  The field in the semiconductor depletion region ($z_{\rm depl}$ in Figure A.4) is $\mathcal{E}_z=\mathcal{E}_{z,\rm{vac}}/\epsilon_{s}$, where $\epsilon_{s}$ is the relative dielectric constant of silicon.  

The electric field $\mathcal{E}_{z}$ during the spatial D$^0$ measurement was estimated using the measured dependence of $U_1$ and $U_2$ on $z$, shown in Figure A.5.C.  Least-square fits on data in Figure A.5.C give vacuum electric fields $-dU_1/dz=(-39\pm16)$ MV/m and $-dU_2/dz=(140\pm15)$ MV/m at peak voltages $U_1=-0.7742 \pm 0.0002$ V and $U_2=-1.0488\pm 0.0005$ V, and the corresponding electric fields $\mathcal{E}_z=\mathcal{E}_{z,\rm{vac}}/\epsilon_{s}$ in the silicon depletion region are $\mathcal{E}_{z,1}=(-3.4\pm1.4)$ MV/m and $\mathcal{E}_{z,2}=(12.3\pm1.3)$ MV/m.  Both the electric field $\mathcal{E}_z$ at $U=-0.85$ V where spatial data was recorded and the flatband voltage where $\mathcal{E}_z=0$ are readily obtained by linear interpolation.  We obtain $\mathcal{E}_z=(-0.3 \pm 1.9)$ MV/m and $U_{FB}=-0.830\pm0.034$ V.  

The inverse decay length $\kappa$ of the vacuum tunneling current $I\propto \exp(-2\kappa z)$ predicted by this electrostatic model was compared against the measured inverse decay length obtained from $dI/dz$ evaluated at $U_{\rm FB}$ in Figure A.2.A.  The dependence of $I$ on $z$ in Figure A.3.A yields $\kappa=1.06\times10^{10}$ 1/m.  In flatband, the tunnel barrier $\Phi_B$ is rectangular and can be approximated as $\Phi_B=\chi+(E_C-E_i)$, where $E_C-E_i\approx0.05$ eV is the neutral donor binding energy and $\chi=4.05$ eV is the electron affinity.  The vacuum tunneling decay parameter predicted for this barrier height\cite{Tersoff:1985bw} is $\kappa=\sqrt{2m_0\Phi_B}/\hbar=1.04\times10^{10}$ 1/m, in excellent agreement with the measurement.  

It can be independently confirmed using $dI/dU$ measurements and extracted values for $\alpha$, and $U_{FB}$, that the lower energy states ($U\approx-0.8$ V) attributed to D$^0$ ground states using $|\Psi(\mathbf{r})|^2$ measurements, are indeed the shallow D$^0$ charge states of donors.  The position of the donor 1 level relative to the sample's Fermi energy $E_F$ is given by\cite{Mol:2013dj} $E_1-E_F = -\alpha e (U_1-U_{FB})$.  Using $U_1$ and $\alpha$ listed in Table 1 and $U_{FB}=-0.83\pm 0.034$ V, we obtain $E_1-E_F \approx 4.9 \pm 1.9$ meV and $E_1-E_F \approx 4.6 \pm 2.5$ meV for donor 1 and donor 2, respectively.  Since the Fermi energy $E_F$ in a degenerately doped reservoir is located approximately at the ionization energy, the ionization energies reported here for subsurface donors are similar to the bulk values for shallow arsenic donors.  

The charging energy of donors 2 and 1 were estimated using the parameters in Table 1.  The expression $E_2-E_1=\frac{1}{2}e(\alpha_1+\alpha_2)(U_2-U_1)$ was employed, which assumes a linear variation in $\alpha$ with voltage between $U_1$ and $U_2$.  Listed in Table 1, $21\pm 2$ meV and $20\pm3$ meV were obtained for donor 1 and donor 2, respectively, considerably less than the $\sim50$ meV charging energy of arsenic in bulk silicon.  Charging energies for subsurface donors of comparable depths have been predicted to be suppressed compared to bulk values using the self-consistent Hartree approach\cite{Rahman:2011cu} in similar, though not directly comparable, geometries.  Since the two-electron state is not electrically neutral, it is expected to be more sensitive to its environment than the neutral state.  We expect that the experimental approach presented here, where electric fields, donor depths, and full spatial probability densities can be determined with great accuracy, will be of utility for studying two-electron states of donors.  Indeed, the spatial structure of the two-electron state has both probability envelope and ellipsoid features in reciprocal space.  Theoretically represented by an overlap between the one-electron and two-electron states\cite{Rontani:2005eb}, the spatial structure of the two-electrons states is however beyond the scope of the present work.  

\begin{figure*}[ht]
\includegraphics{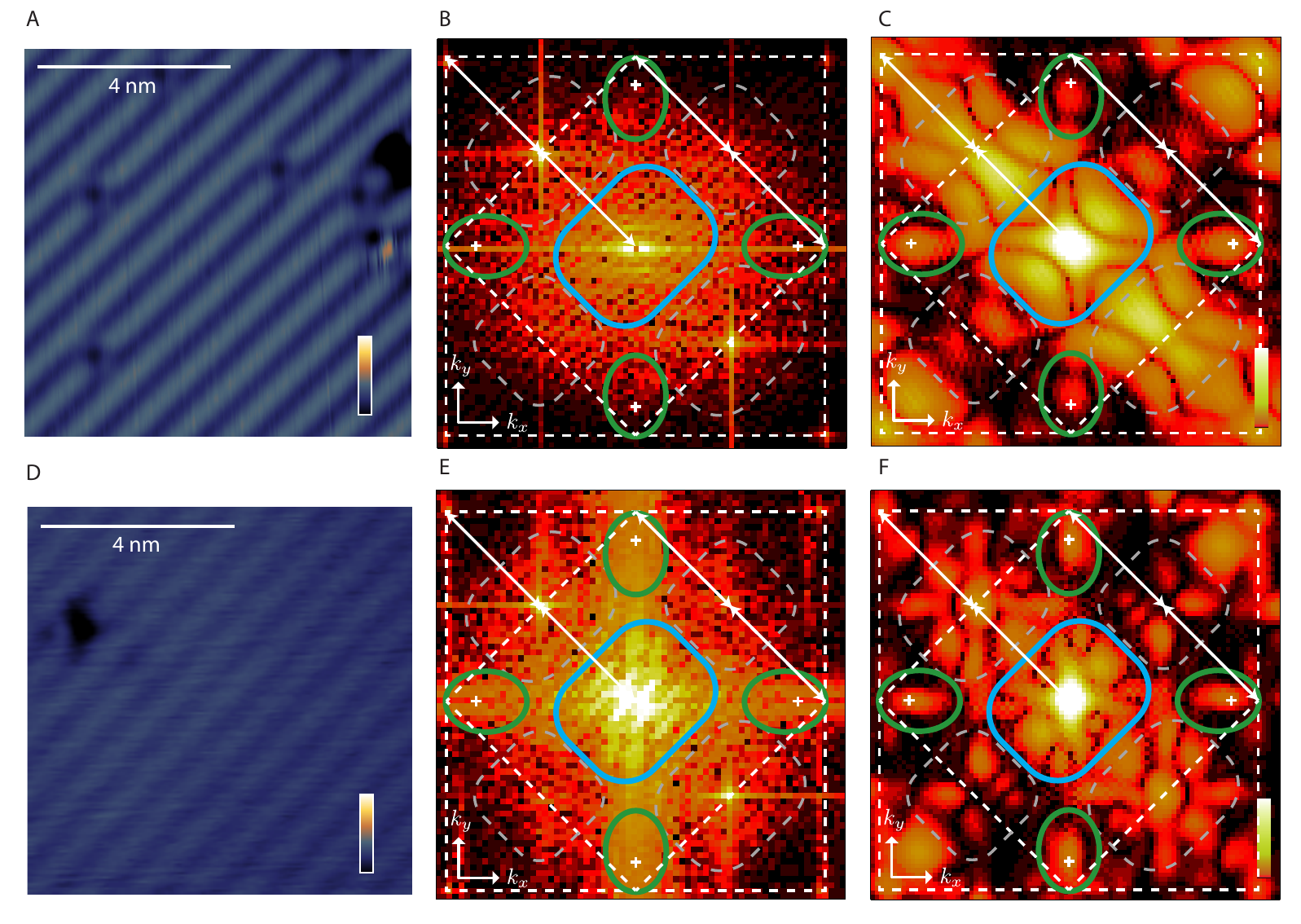}
\begin{flushleft}
{Fig. A.6.A  Closeup 8 nm $\times$ 8 nm of topography centered over donor 1 from the main text.  Scale bars are (0,100 pm).  B. Fourier transform of topography $z(x,y)$ in 20 nm $\times$ 20 nm frame for donor 1, and overlay reciprocal space map from main text.  Corners of the outer rectangle are reciprocal lattice vectors $2\pi/a_0(p,q)$ where $p=\pm 1$ and $q=\pm1$.  C.  Fourier transform of donor 1 ground state from main text.  D.  Closeup 8 nm $\times$ 8 nm of topography over donor 2 from the main text.  Scale bars are (0,100 pm).  E.  Fourier transform of topography $z(x,y)$ in $20\times20$ nm frame for donor 2, and overlay reciprocal space map from main text.  F.  Fourier transform of donor 2 ground state from main text.  }
\end{flushleft}
\end{figure*}

\subsection{Reconstruction-induced features}

In this section we discuss calibration of reciprocal lattice vector positions, carried out using Fourier transforms of measured topographies, as well as the appearance and origin of reconstruction-induced features in both the topographies and measured quantum states, and the associated sampling requirements to faithfully represent them in the measurements. The topography $z(x,y)$ measured simultaneously with donor 1 (donor 2) in the main text is given in Figure A.6A (A.6D). 

Peaks at reciprocal lattice frequencies in the the corresponding Fourier transform for donor 1 (donor 2) given in Figure A.6B (A.6E) were employed to carry out a fine calibration of the coordinate system.  Results are shown after calibration correcting a trivial rotational misalignment of the cleaved 10 mm $\times$ 3 sample within the sample plate, and a small tip drift\cite{Rahe:2010gy} ($< 0.1$ nm/hour) of the 4.2 K LT-STM during the $\sim 30$ minute measurement of each donor.  

The result is the alignment of reciprocal lattice vector positions $2\pi/a_0(\pm1,\pm1)$ in the measured topography with spots corresponding to the cubic lattice constant $a_0=0.543$ nm of the silicon surface.  Peaks at $2\pi/a_0(\pm 1/2, \mp 1/2)$ originate from the $2\times1$ surface reconstruction.  The latter, created by symmetric dimerization of the hydrogen terminated silicon surface\cite{Craig:1990ge}, are found at a displacement $\Delta \mathbf{G} = 2\pi/a_0(\mp1/2, \pm1/2)$ relative to reciprocal lattice vectors positions.  Features centered at the same coordinates were observed in the Fourier transform of the orbital probability density of donor 1 (Figure 2A and Figure A6.C) and donor 2 (Figure A.6F), a displacement of $\Delta \mathbf{G}$ from probability envelope features at $\mathbf{k}=2\pi/a_0(\pm1,\pm1)$ and $k=0$.  Similarly, the structure found at $2\pi/a_0(\pm1/2, \pm1/2)$ for donor 1 (Figure A4.C) and donor 2 (Figure A4.F) is related to the four symmetric ellipsoids centered around $2\pi/a_0(0,\pm1)$ and $2\pi/a_0(\pm1,0)$, as evidenced by the $\Delta \mathbf{G} = 2\pi/a_0(\mp1/2, \pm1/2)$ displacement relative to these structures.  

For the sake of simplicity, the first Brillouin zone of measured and calculated Fourier representations are shown.  However, measurements reveal nonzero components with spatial frequencies to at least $2(2\pi/a_0)$.  The Nyquist sampling theorem therefore dictates a minimum spatial sample frequency exceeding $4(2\pi/a_0)$.  Using our single quantum state imaging method we could comfortably image single states with a spatial resolution of $1024 \times 1024$ pixels (for donor 1) and $512 \times 512$ pixels (for donor 2), in less than 30 minutes.  The corresponding maximum spatial frequencies represented in the $L \times L=20$ nm $\times$ 20 nm frames, providing the desired resolution $2\pi/L\approx(1/50)2\pi/a_0$, are $\sim 12(2\pi/a_0)$ (donor 1) and $\sim 6(2\pi/a_0)$ (donor 2).  

\subsection{Valley Population}

\begin{figure*}[ht]
\includegraphics{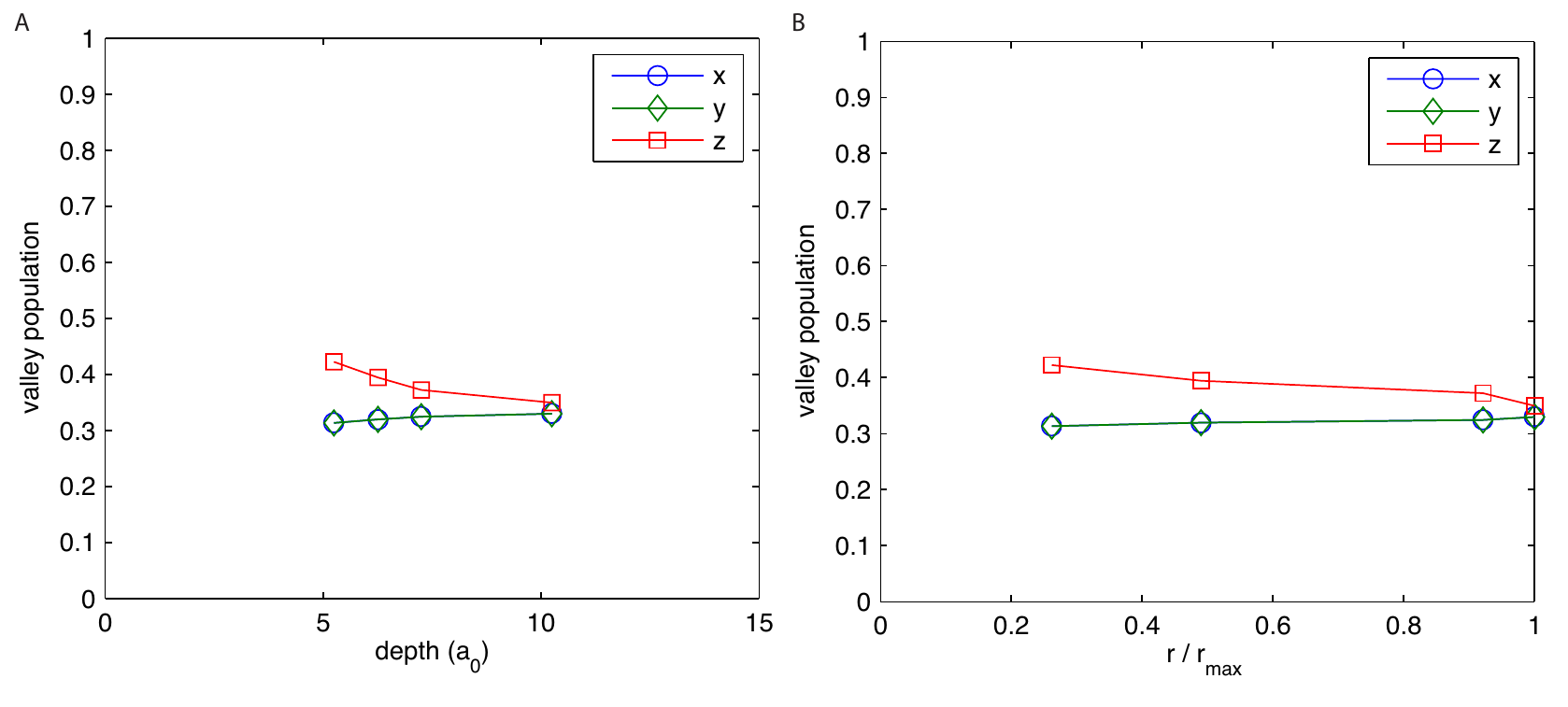}
\begin{flushleft}
{Fig. A.7. Relative distribution of probability amplitudes of donor wavefunction between $x$, $y$, and $z$ valleys as a function of (A) donor depth, and as a function of (B) peak ratio.  }
\end{flushleft}
\end{figure*}

The absolute valley population for donor 1 was estimated by comparing the ratio of peaks at $\mathbf{k}=0$ and $\mathbf{k}=0.15(2\pi/a_0)(\pm 1, \pm 1)$ in measurements to calculations.  In the main text we found that the ratio matched calculations for donor depths in zero field between $6.25a_0$ and $7.25a_0$.  The corresponding absolute coherent valley population of the measured donor-bound state was estimated from the absolute coherent valley population calculated for depths $6.25a_0$ and $7.25a_0$.  The latter was determined by a three-dimensional Fourier decomposition of the tight-binding wavefunction $\Psi(\mathbf{r})=\sum_{i,j}c_{ij}\phi_i(\mathbf{r}-\mathbf{R}_j)$.   Writing $\phi(r)=\sum_k \phi_i(\mathbf{k}) \exp(\mathbf{k}\cdot \mathbf{r})$ and $c_{ij}=c_i(\mathbf{R}_j)$ we obtain 
\begin{equation}
\Psi(\mathbf{r}) = \sum_k \exp(i\mathbf{k}\cdot \mathbf{r})\Big[\sum_i \phi_i(\mathbf{k})\sum_j c_i(\mathbf{R}_j) \exp(-i\mathbf{k}\cdot\mathbf{R}_j)\Big].
\end{equation}
The term in block parenthesis is readily recognized as the Fourier component $\Psi(\mathbf{k})$ while the sum over $j$ is a discrete Fourier transform, which was evaluated on a grid.  The relative contribution of $x$, $y$ and $z$ valleys were obtained by integrating the three-dimensional distribution of valleys in $\Psi(\mathbf{k})$ obtained by this method.  Results for $x$, $y$, and $z$ valleys are shown in Figure A.7 as a function of donor depth (Figure A.7.A) and peak ratio (Figure A.7.B).  

\subsection{Real-space representation of ellipsoid Fourier feature}

The expression for the real-space representation of the ellipsoid valley interference pattern in the main text was derived from the six-valley donor envelope function representation introduced by Kohn and Luttinger\cite{Kohn:1955kc},
\begin{equation}
\Psi^i(\mathbf{r})=\sum_\mu \alpha^i_{\mu}F_\mu(\mathbf{r})\phi_{\mathbf{k}_\mu}(\mathbf{r}),
\end{equation}
where $\mu=1\dots6$ denotes the six conduction band minima in silicon, $\alpha^i_\mu$ denotes the valley quantum number describing the coherent valley population of the minimum, $F_\mu(\mathbf{r})$ is the envelope function, $\phi_{\mathbf{k}_\mu}(\mathbf{r}) = \exp(i\mathbf{k}_\mu\cdot\mathbf{r})u_{\mathbf{k}_\mu}(\mathbf{r})$ is the Bloch function for the band minimum, and $u_{\mathbf{k}_\mu}(\mathbf{r})=\sum_\mathbf{G}A_{\mathbf{k},\mathbf{G}}\exp(i\mathbf{G}\cdot\mathbf{r})$ is a lattice periodic function over the reciprocal lattice vectors $\mathbf{G}$ of the silicon crystal.  

Supported by atomistic calculations, the measured subsurface donor ground state was found to have a bulk-like orbital structure and valley configuration.  Consequently, we take $\boldsymbol{\alpha}^1=1/\sqrt{6}(1,1,1,1,1,1)$.  As discussed in the main text, the ellipsoid features arise from products $z$ and $x$ (or $y$) valleys in $|\Psi(\mathbf{r})|^2$.  Selectively expanding those terms using $\Psi(\mathbf{r})$ from above and assuming the ion position is located at $x=y=z=0$, it is easy to show that their contribution to the probability density evaluated at $(x,y,z_0)$ is given by
\begin{align}
P(x,y,z_0)=CF_z(x,y,z_0) \times \\ [F_x(x,y,z_0)\cos(k_\mu x) + F_y(x,y,z_0)\cos(k_\mu y)]
\end{align}
where $C=8\cos(k_\mu z_0)\sum_\mathbf{G} |A_{\mathbf{k}_\mu,\mathbf{G}}|^2$.  This expression has a global maximum at $x=y=0$ when $\cos(k_\mu z_0) > 0$ and a global minimum in the same location when $\cos(k_\mu z_0) < 0$.  The ion position indicated in Figure 4 in the main text can be determined to high accuracy $\delta_x,\delta y \ll 0.85(2\pi/a_0)$ by examining the global minimum in the real-space representation of the valley oscillation.  

\subsection{Theory}
\begin{figure*}[ht]
\includegraphics{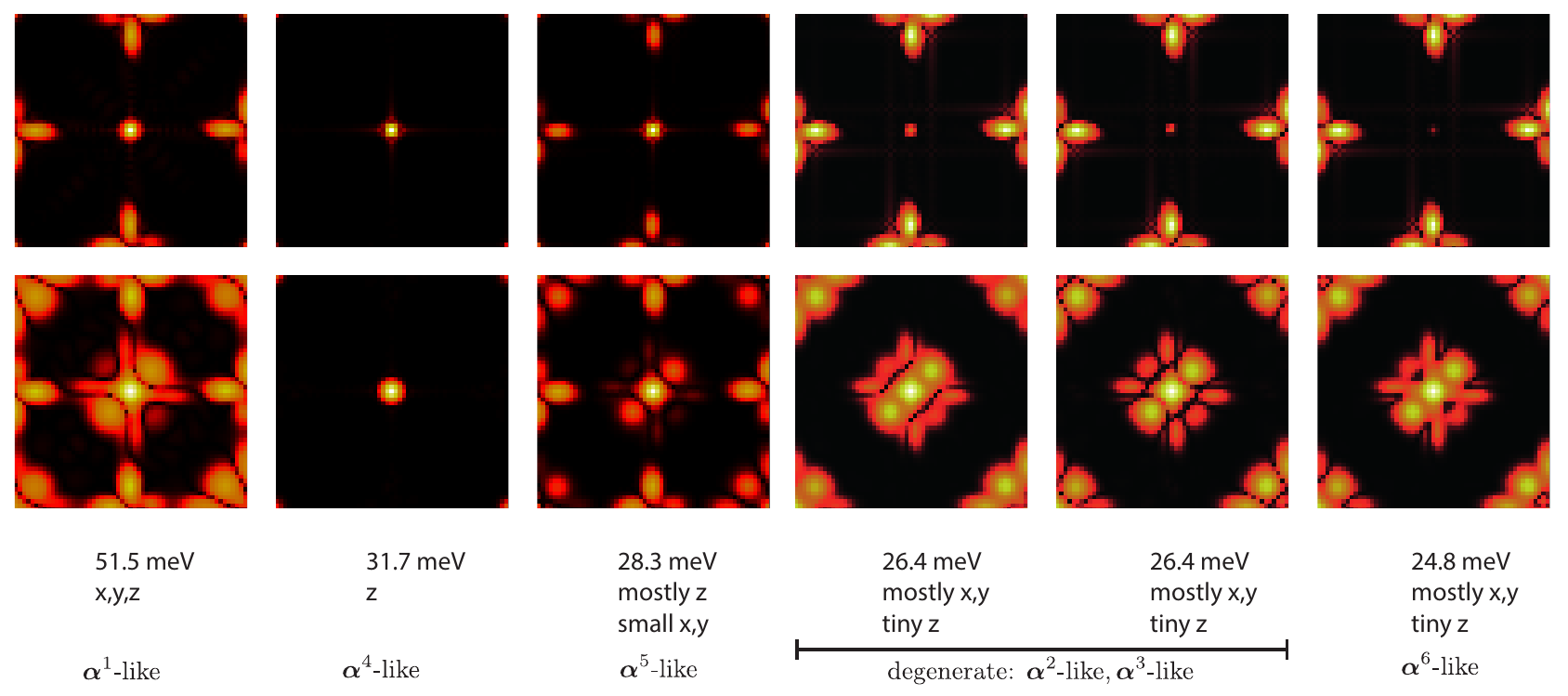}
\begin{flushleft}
{Fig. A.8. Fourier transform of tight-binding calculations for subsurface donor $6.25a_0\approx2.8$ nm below hydrogen-terminated silicon surface, for region $k_x=[-2\pi/a_0,2\pi/a_0]$ and $k_y=[-2\pi/a_0,2\pi/a_0]$.  First row is $\Psi(\mathbf{r})$.  Second row is $|\Psi(\mathbf{r})|^2$.  Columns $i=1\dots6$ are first 6 valley eigenstates.  Energies, and dominant contribution to valley configuration are listed in each column.  Corresponding bulk-like valley configuration is given.  }
\end{flushleft}
\end{figure*}

Atomistic predictions of the electron wavefunctions $\Psi_i(\mathbf{r})$ and probability densities $|\Psi_i(\mathbf{r})|^2$ of donor-bound states were obtained by empirical sp$^3$d$^5$s$^*$ tight-binding.  Both were evaluated in the evanescent (vacuum) tail of the wavefunction $\Psi_1(\mathbf{r})=\sum_{j,k} c^1_{j,k}\varphi_j(\mathbf{r}-\mathbf{R}_k)$, where $c^1_{jk}$ is the tight-binding representation of the ground state wavefunction, $\varphi_j(\mathbf{r})$ are s, p, and d Slater-type orbitals for silicon\cite{Nielsen:2012wt}, and $\mathbf{R}_k$ are positions of silicon atoms in the crystal.  Evanescent (vacuum) tails of $\Psi(\mathbf{r})$ and $|\Psi(\mathbf{r})|^2$ were evaluated for a tip orbital position $\delta z=0.45$ nm above the last atomic plane, assuming an s-wave tip\cite{Chen:1990cr}.  

Coefficients $c^1_{jk}$ above were obtained by diagonalization of the tight binding Hamiltonian of $\sim1.4$ million silicon atoms. A single arsenic atom was modeled as a Coulomb potential with on-site orbital energy corrections\cite{Lansbergen:2008bs} successfully reproducing the measured donor energy spectrum including the valley-orbit splitting\cite{Ramdas:2000it}. The influence of displacement of surface atoms associated with the $2\times1$ surface reconstruction on the tight binding Hamiltonian was computed by a generalization of Harrison's scaling law\cite{Boykin:2002fu}, assuming symmetric dimerization atomic displacements calculated elsewhere\cite{Craig:1990ge}. These models have been combined with the valence force field Keating model to describe atomistic strain relaxation in multi-million atom quantum dots in excellent agreement with experiments\cite{Klimeck:2007gs}. 

The full tight-binding Hamiltonian of silicon, the arsenic donor, and the hydrogen passivated surface was solved by a parallel Lanczos eigensolver to obtain the lowest energy donor eigenstates.  The computations are performed using the atomistic tight-binding tool NEMO-3D utilizing nanohub.org resources, and requires about 5 hours on 40 processors to obtain 10 wavefunctions. A detailed description of this full-band non-perturbative method can be found in Refs \onlinecite{Rahman:2007ev,Klimeck:2007gs}.  

\subsection{Calculated single-electron excited states of subsurface and bulk donor}

Further details of tight-binding calculations of subsurface donor states are presented in this section.  Two-dimensional Fourier transforms of wavefunctions $\Psi(x,y,z_0)$ and probability densities $|\Psi(x,y,z_0)|^2$ of the first six single-electron (D$^0$) eigenstates of a subsurface donor, calculated by tight binding, are given in Figure A.8.  The binding energy relative to the bottom of the bulk silicon conduction band is given, along with the dominant valley contribution and the corresponding valley quantum numbers.  

In particular, the predicted binding energy of the subsurface arsenic donor (51.5 meV) is only slightly smaller than that of the bulk arsenic donor (53.4 meV).  Similarly, the valley-orbit splitting of the subsurface donor is $19.8$ meV compared to the 21.1 meV splitting predicted for the bulk donor.  The first excited state is $\boldsymbol{\alpha}^4$-like and the second is $\boldsymbol{\alpha}^5$-like.  The third and fourth states are degenerate and are both linear combinations of $\boldsymbol{\alpha}^2$ and $\boldsymbol{\alpha}^3$-like states, while the highest energy state in the manifold is $\boldsymbol{\alpha}^6$-like. Recall for a donor in bulk silicon, the singlet has $\boldsymbol{\alpha}^1=6^{-1/2}[1,1,1,1,1,1]$, the triplet has $\boldsymbol{\alpha}^2=2^{-1/2}[1,-1,0,0,0,0]$, $\boldsymbol{\alpha}^3=2^{-1/2}[0,0,1,-1,0,0]$, and $\boldsymbol{\alpha}^4=2^{-1/2}[0,0,0,0,1,-1]$, and the doublet has $\boldsymbol{\alpha}^5=12^{-1/2}[-1,-1,-1,-1,2,2]$ and $\boldsymbol{\alpha}^6=2^{-1}[1,1,-1,-1,0,0]$.

\end{document}